\documentclass[twocolumn]{paper}
\usepackage{graphicx}
\usepackage{psfrag}

\author{M. Loewe
         and
         C. Villavicencio
            \\{\small Facultad de F\'{\i}sica,
 Pontificia Universidad Cat\'olica de Chile, Casilla 306, Santiago 22, Chile}}
\title{Thermal Pions and Isospin Chemical Potential Effects}

\begin{document}
\maketitle
\begin{abstract}
The density corrections, in terms of the isospin chemical
potential $\mu_I$, to the mass of the pions are investigated in
the framework of the $SU(2)$ low energy effective chiral invariant
lagrangian. As a function of temperature and $\mu_I =0$, the mass
remains quite stable,  starting to grow for very high
 values of $T$,  confirming previous results. However,
  the dependence for  a non-vanishing chemical potential turns out to be much more dramatic.
In particular, there are interesting corrections to the mass when
both effects (temperature and chemical potential) are
simultaneously present. At zero temperature the $\pi ^{\pm}$
 should condensate when $\mu _{I} = \mp m_{\pi }$. This is not
 longer valid anymore at finite $T$. The mass of the $\pi_0$
 acquires also a non trivial dependence on $\mu_I$ at finite $T$.
\end{abstract}

\bigskip

Pions play a special role in the dynamics of hot hadronic matter
since they are the lightest hadrons. Therefore, it is quite
important to understand not only the temperature dependence of the
pion's Green functions but also their behavior as function of
density. The dependence of the pion mass (and width) on
temperature $m_{\pi}(T)$ has been studied in a variety of
frameworks, such as thermal QCD-Sum Rules \cite{DFL}, Chiral
Perturbation Theory (low temperature expansion) \cite{GL}, the
Linear Sigma Model \cite{LCL}, the Mean Field Approximation
\cite{Bar1}, the Virial Expansion \cite{Schenk}. In fact the
properties of pion propagation at finite temperature have been
calculated at two loops in the frame of chiral perturbation theory
\cite{Schenk2}. There seems to be a reasonable agreement that
$m_{\pi}(T)$ is essentially independent of $T$, except possibly
near the critical temperature $T_{c}$ where $m_{\pi}(T)$ increases
with $T$.

Let us proceed in the frame of the $SU(2)$ chiral perturbation
theory. The most general chiral invariant expression for a
QCD-extended lagrangian, \cite{GL2,pich} under the presence of
external hermitian-matrix auxiliary fields has the form
\begin{eqnarray}
{\cal L}_{\tiny QCD}(s,p,v_\mu,a_\mu)   &=&  {\cal L}_{\tiny QCD}^0\nonumber\\
                                        &+& \bar q \gamma^\mu (v_\mu +\gamma_5 a_\mu)q\nonumber\\
                                        &-& \bar q(s-i\gamma_5
                                        p)q,
\end{eqnarray}
where  $v_\mu$, $a_\mu$, $s$ and $p$ are vectorial, axial, scalar
and pseudoscalar fields. The vector current  is given by
\begin{equation}
J_\mu^a=\bar q \gamma_\mu \tau^aq.
\end{equation}

When  $v,a,p=0$ and $s=M$, being $M$ the mass matrix, we obtain
the usual $QCD$ Lagrangian. The effective action with finite
isospin chemical potential is given by
\begin{eqnarray}
{\cal L}_{\tiny QCD}^I    &=& {\cal L}_{\tiny QCD}(M,0,0,0)+\mu^a u^\mu  J^a_\mu\nonumber\\
                    &=& {\cal L}_{\tiny QCD}(M,0,\mu u_\mu,0) \label{eq:accion}
\end{eqnarray}
where $\mu^a=(0,0,\mu_I)$ is the third isospin component,
$\mu=\mu^a\tau^a$ and $u_\mu$ is the 4-velocity between the
observer and the thermal heat bath. This is required to describe
in a covariant way this system, where the Lorentz invariance is
broken since the thermal heath bath represents a privileged frame
of reference.

Proceeding in the same way in the low-energy description, where
only the pion degrees of freedom are relevant, let us consider the
most general chiral invariant lagrangian. This lagrangian is
ordered according to a series in powers of the external momentum.
We will start with the ${\cal O}(p^2)$ chiral lagrangian
\begin{equation}
{\cal L}_2=\frac{f^2}{4}Tr\left[(D_\mu U)^\dag D^\mu U+U^\dag \chi
+ \chi^\dag U\right]
\end{equation}
with
\begin{eqnarray}
 D_\mu U &=& \partial_\mu U-i[v_\mu, U]-i\{ a_\mu, U\}\nonumber\\
 \chi &=& 2B_0(s+ip)\nonumber\\
 U &=& e^{i2\pi^a\tau^a/f}\label{eq:fields}
\end{eqnarray}
\noindent $B_{0}$ in the previous equation is an arbitrary
constant which will be fixed when the mass is identified setting
$(m_{u} + m_{d})B_{0} =m^{2}$.
 The most general ${\cal O}(p^4)$ chiral lagrangian has the form ${\cal L}_4=\sum\alpha_i{\cal L}_{(\alpha_i)}$
  being the  SU(3) lagrangian
 related to the SU(2) one \cite{holstein,Schenk2}.
 We will work with the general SU(3) ${\cal O}(p^4)$ lagrangian.
The effective action with finite chemical potential in terms of
pion degrees of freedom has the same form as eq.\ref{eq:accion},
where the different external fields are defined in
eq.\ref{eq:fields}. In this paper we will consider one loop
corrections, up to the fourth order in the fields, to the
lagrangian ${\cal L}_2$ and the free part, i.e the tree level part
of ${\cal L}_4$ with renormalized fields. This procedure is
standard, \cite{GL2,holstein}.
 The interacting part ${\cal L}_{4}$ involves higher powers in the momentum  of the pion fields.
  The constants $\alpha _{i}$ present in ${\cal L}_{4}$ are known  from decay and scattering measurements.
\begin{eqnarray}
    {\cal L}_{2,2}^I &=&
        \frac{1}{2}\left[\left(\partial\pi_0\right)^2-m^2\pi_0^2\right]
        +\left|\partial_I\pi\right|^2-m^2\left|\pi\right|^2\nonumber\\
{\cal L}_{2,4}^I &=& \frac{1}{4!}\frac{m^2}{f^2}\pi_0^4 +
\frac{1}{6f^2}\left[
-4\left|\partial_I\pi\right|^2\left|\pi\right|^2
\right.   \nonumber\\
               & &  \qquad\left.  +\left(\partial\left|\pi\right|^2\right)^2 + m^2\left(\left| \pi\right|^2\right)^2\right]\nonumber\\
                & & +\frac{1}{6f^2}\left[ -2\left|\partial_I\pi\right|^2\pi_0^2
                                            -2\left(\partial\pi_0\right)^2\left|\pi\right|^2
                                            \right.\nonumber\\
                & &  \qquad \quad    \left.   +\partial\pi_0^2\cdot\partial\left|\pi\right|^2
                +m^2\pi_0^2\left|\pi\right|^2\right]\nonumber\\
{\cal L}^{I}_{4,2}   &=&
\frac{m^2}{32\pi^2f^2}\left\{a\left|\partial_I\pi\right|^2
                                         -m^2b\left|\pi\right|^2\right.\nonumber\\
                    & &       +  \left.\frac{1}{2}a\left(\partial\pi_0\right)^2
                        -\frac{1}{2}m^2(b+\epsilon_{ud}^2 c)\pi_0^2\right\}
\label{eq:a,b}
\end{eqnarray}
with
\begin{eqnarray}
a   &=& 32\pi^2(16\alpha_4+8\alpha_5)\nonumber\\
b   &=& 32\pi^2(32\alpha_6+16\alpha_8)\nonumber\\
c   &=& 32\pi^2(32\alpha_7+16\alpha_8)
\end{eqnarray}
where the subindexes ($i,j$) in the lagrangian denote the order in
powers of momentum and  fields, respectively, and
 $\partial_I\equiv
\partial+i\mu_Iu$. This definition of the covariant derivative is
natural according to our previous comments, since we know
\cite{weldon,actor}  that the chemical potential is introduced as
the zero component of an external ``gauge" field. In the previous
expression, $|\pi|^2$ means $\pi^+\pi^- = \pi\pi^*$, and
$|\partial_I\pi|^2=(\partial_I \pi)^*(\partial_I \pi)$. $\epsilon
_{ud}^2  = (m_u-m_d)^{2}/(m_u + m_d)^{2}$ will be neglected
because it only shifts in a small quantity the neutral pion mass
and we are interested in the thermal and density evolution of the
masses.

For renormalizing with counterterms we introduce the following
decomposition
\begin{eqnarray}
    {\cal L}_{eff}  &=& {\cal L}_{2,2}^I+ {\cal L}_{2,4}^{Ir}+{\cal L}^{Ir}_{4,2}\nonumber\\
     {\cal L}_{2,2}^I   &=& {\cal L}^{Ir}_{2,2}+\delta{\cal L,}
\end{eqnarray}
where the $r$ index denote the lagrangian with renormalized
fields. These changes are related to higher loop corrections.

Setting $\pi_0=\sqrt{Z_0}\pi_0^r$ and
$\pi_\pm=\sqrt{Z_\pm}\pi_\pm^r$ in ${\cal L}_{2,2}$, we have
\begin{eqnarray}
\delta{\cal L} &=&
\frac{1}{2}\delta_{Z_o}\left[(\partial\pi_0^r)^2
                    -m^2(\pi_0^r)^2\right]\nonumber\\
                 & &   +\delta_{Z_\pm}\left[\left|\partial_I\pi^r\right|^2
                    -m^2\left|\pi^r\right|^2\right]
\end{eqnarray}
with $\delta_{Z_i}=Z_i-1$.

First, let us consider the thermal and density corrections to the
propagator, the two-point function pion correlator. Since our
calculation will be at the one loop level, we do not need the full
formalism of thermo field dynamics, including thermal ghosts and,
therefore, matrix propagators. The propagator in momentum space
$D(k;T,\mu_I,m^2)=D(k;\mu_I,m^2)+D_T(k,\mu_I,m^2)$ for charged
pions at the tree level will be given by an extension, for a
non-vanishing chemical potential, of the well known Dolan-Jackiw
propagators for scalar fields \cite{weldon}. Note that since there
is no chemical potential associated to the neutral pion, the
thermal propagator $D_{0}$ will be the usual one
$D_0(k;T,\mu_I,m^2)=D(k,T,0,m^2)$ with
\begin{eqnarray}
D(k;\mu_I,m^2)  &=& \frac{i}{k_+^2-m^2+i\epsilon}\nonumber\\
D_T(k;\mu_I,m^2)&=& 2\pi n_B(|k\cdot u|)\delta(k_+^2-m^2)
\end{eqnarray}
where $k_\pm\equiv k\mp\mu_Iu$ and $n_B(x)=(e^{x/T}-1)^{-1}$ is
the Bose-Einstein factor.

\bigskip
\noindent

 We will use the $\overline{MS}$-scheme, and we renormalize as usual at $T=0$, since the thermal corrections
are finite. The self energy for charged pions including the
counterterms has the form
\begin{equation}
\Sigma(p)  =
[A-\delta_{Z_\pm}]p_+^2-[A'-\delta_{Z_\pm}]m^2+A''u\cdot p_+
\end{equation}

 Our
prescription to fix the counterterm $\delta_{Z_\pm}$ is to impose
that $\Sigma$ does not depend on $p^2$, so, $\delta_{Z_\pm}=A$. In
this way, the inverse propagator will take the form
$i[D(p;\mu_I,m^2+\Sigma)]^{-1}=p^2+Cp_0+C'$ in the frame where the
heath bath is at rest ($u=(1,0,0,0)$).  The $C$ term will give two
solutions to $i[D(p_0,{\bf p}=0;\mu_I,m^2+\Sigma)]^{-1}=0$ that we
identify as $m_{\pi^+}$ and $m_{\pi^-}$. The
$\alpha_i=\alpha_i^r(\Lambda)-\frac{\gamma_i}{32\pi^2}\left[\frac{2}{d-4}-\ln
4\pi +\gamma -1\right] $ absorb the divergences. The $\gamma_i$
terms are tabulated \cite{GL2,holstein}.  The well known result
for $T=\mu_I=0$
\begin{equation}
m_\pi^2=m^2\left[
1-g\left\{32\pi^2(a^r-b^r)+\ln\frac{m_\pi^2}{\Lambda^2}\right\}\right]
\end{equation}
is identified with the physical mass with $\Lambda$ a scale
factor. $g=m_\pi^2/32\pi^2f_\pi^2$ is the perturbative term that
fixes  the scale of energies in the theory (for energies below
$32\pi^2f_\pi^2$) so we neglect the ${\cal O}(g^2)$ factors. This
allows us to set $m \approx m_\pi$ in all radiative corrections
(and also $f \approx f_\pi$). The procedure is the same for
$m_{\pi^0}$

It is important to remark that radiative corrections will leave a
dependence on the chemical potential for the pion mass only for
finite values of temperature. In a strict sense, this procedure
does not allow us to say nothing new for an eventual chemical
potential dependence of the masses at $T=0$ (cold matter) which is
already included in ${\cal L}_2$. In this case, $T=0$, we have to
follow the usual procedure, \cite{son,toublan}, of computing the
minimum of the effective potential in ${\cal L}_2$ when the
chemical potential is taken into account, without considering
radiative corrections. This enables to identify a phase structure
where a non trivial vacuum appears for higher values of $\mu
_{I}$, characterized by the appearance of a condensate $\langle
\pi ^{-}\rangle$. (The opposite occurs for negative values of the
chemical potential, where  the vacuum state is a condensate
$\langle \pi ^{+} \rangle $ for $|\mu _{I}| > m_{\pi }$). At $T=0$
when $\mu _{I} = m_{\pi }$, the mass of $\pi ^{-} $ vanishes.

For finite $T$ and $\mu_I$, we find the following expression for
the masses
\begin{eqnarray}
m_{\pi^\pm}(T,\mu_I)\!\! &=&  \!\!  m_\pi\left[1+2g
I_{(0)}\pm\left(\frac{\mu_I}{m_\pi}-8g J\right)\right]\nonumber\\
m_{\pi^0}(T,\mu_I)\!\! &=&\!\!
m_\pi\left[1+2g\left(2I_{(\mu_I)}-I_{(0)}\right)\right]
\end{eqnarray}
with
\begin{eqnarray}
I_{(\mu_I)}\!\!\!\!\!&=& \int\!\!\!
    \frac{d^4k}{(2\pi m_\pi)^2}n_B(|k_0+\mu_I|)\delta(k^2-m_\pi^2)\nonumber\\
  J \!\!\!\!\!&=&\!\!\!\!\int\!\!\! \frac{d^4kk_0}{(2\pi
  m_\pi)^2}n_B(|k_0+\mu_I|)\delta(k^2-m_\pi^2)
\end{eqnarray}

Note that our convention for the chemical potential is contrary to
the one adopted in the paper by Kogut and Toublan \cite {toublan},
 who extended previous results by Son and Stephanov \cite{son}.

If the chemical potential of the charged pions vanishes, i.e for
symmetric matter,  at finite T we get the well known result for
$m_{\pi }(T)$ due to chiral perturbation theory \cite{GL}, see
also \cite{LCL}.
 However, due to radiative corrections to the neutral pion propagator,
  its mass will acquire a non trivial chemical potential dependence for finite values of temperature.
   In the approach where the minimum of the effective potential is calculated (for finite $\mu _{I}$ and $T=0$),
    the mass of the neutral pion remains constant.

\begin{figure}
\fbox{\begin{minipage}{7.8cm}
\psfrag{z}{\bf\large $m_{\pi^0}/m_\pi$} \psfrag{xxxx}{$T/m_\pi$}
\psfrag{y}{$\mu_I/m_\pi$}
\includegraphics[scale=1]{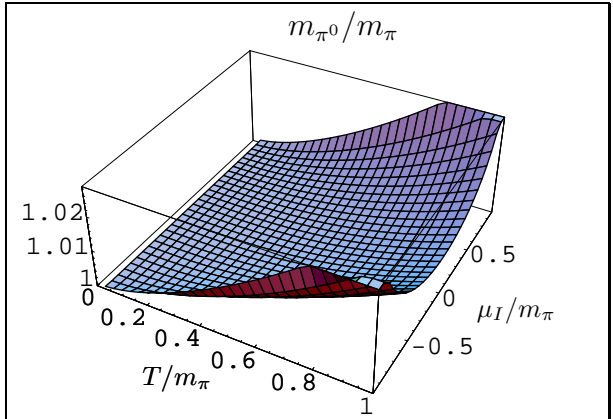}
\caption{$m_{\pi^0}$ as function of $T$ and $\mu_I$ in units of
$m_\pi$ } \label{fig1}
\end{minipage}}
\end{figure}

We show in Fig.\ref{fig1} a tridimensional picture for the
behavior of the mass of the neutral pion.
 Note that when $\mu_I=0$, $m_{\pi_0}(T)=m_{\pi^\pm}(T)$.

\begin{figure}
\fbox{\begin{minipage}{7.8cm}
 \psfrag{x}{$\frac{\mu_I}{m_\pi}$}
\psfrag{y}{$m_{\pi^+}/m_\pi$} \psfrag{T10000}{\small $T=0$}
\psfrag{T20000}{\small $T=0.5m_\pi$} \psfrag{T30000}{\small
$T=m_\pi$} \psfrag{T40000}{\small $T=1.5m_\pi$}
\includegraphics[scale=1]{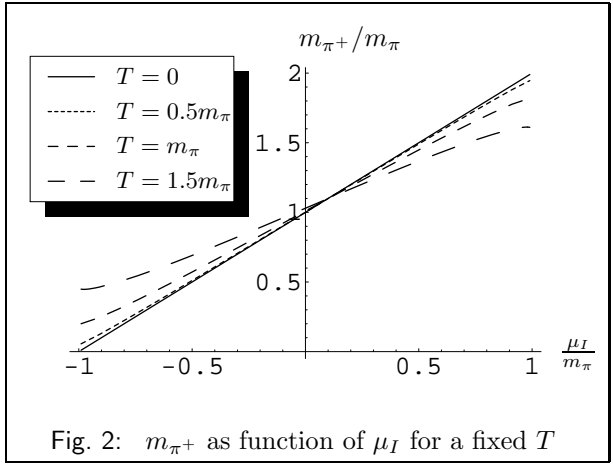}
 \caption{ $m_{\pi^+}$ as function of  $\mu_I$ for a fixed $T$ }
 \label{fig2}
 \end{minipage}}
\end{figure}

From Fig.\ref{fig2} we see that at zero temperature, we agree with
the usual prediction, $m_{\pi }^{+} = m_{\pi } + \mu _{I}$. In
fact, at zero temperature the $\pi ^{+}$ should condensate when
$\mu _{I} = -m_{\pi }$ (the inverse situation occurs for $\pi
^{-}$).
 Now, this situation changes if temperature starts to grow. The
 condensation point disappear at $\mu_I=-m_\pi$; in $\mu_I=m_\pi$ the mass start to decrease
 . For small $T$ (for example inside an neutron star),
  we cannot see such effect.

In order to explore the region where $|\mu _{I}| > m_{\pi}$,
associated to a new phase where the condensates occur,
  we need to redefine our fields as fluctuations around the configuration corresponding to a minima
  of the effective potential in ${\cal L}_{2}$.\\

\noindent
 {\bf Acknowledgements:}   The work of  M.L. has been
supported
 by Fondecyt (Chile)
under grant No.1010976. C.V. acknowledges support from a Conicyt
Ph.D fellowship (Beca Apoyo Tesis Doctoral). The authors thank
Hernan Castillo for helpful discussions and for technical advise
concerning numerical questions.

\end{document}